\documentclass[12pt,prd,superscriptaddress,aps,nofootinbib,floats,floatfix,amsmath,amssymb,secnumarabic]{revtex4}
\usepackage{amsmath}
\usepackage{amsfonts}
\usepackage{amssymb}
\usepackage{latexsym}
\usepackage{graphicx}
\usepackage[english]{babel}

\newcommand{\be}{\begin{equation}}
\newcommand{\ee}{\end{equation}}
\newcommand{\ba}{\begin{array}}
\newcommand{\ea}{\end{array}}
\newcommand{\bqa}{\begin{eqnarray}}
\newcommand{\eqa}{\end{eqnarray}}
\renewcommand{\d}{\mathrm{d}}

\newcommand{\e}[1]{e^{#1}}

\begin{document}

\title{Non-relativistic  Supersymmetry}

\author{Wei Xue\footnote{xuewei@hep.physics.mcgill.ca}}
\affiliation{Department of Physics, McGill University,
3600 Rue University, Montr\'eal, Qu\'ebec, Canada H3A 2T8}

\begin{abstract}

We construct an $N=1$ supersymmetric gauge theory from $z=3$ Lifshitz field theory. 
By modifying the supersymmetry (susy) algebra based on the spacetime symmetry   
$SO(3)$ $\times$ scaling symmetry, we get a supersymmetric Lagrangian with 
scalar, fermion and gauge fields, all of whom have the same 
limiting speed. This solves some naturalness problems of the original Lifshitz
theory which is characterized by Lorentz symmetry violation. In order that the 
susy  breaking does not introduce any disastrous terms into the theory, the
susy breaking scale is required to be smaller than the scale of Lorentz 
symmetry violation.

\end{abstract}
 \maketitle

\newpage

\section{Introduction}

Theory and experiments have been exploring possible  violations of the Lorentz symmetry 
for decades. On the other hand,  Lorentz symmetry has been very successful in 
the Standard Model of Particle Physics and in General Relativity. Low energy Lorentz 
violation possibly originates from high energy physics, such as quantum gravity and 
string theory. A well-known example is that non-commutative geometry in string theory due to 
a non-zero tensor $B_{\mu \nu}$ on a D-brane leads to Lorentz 
symmetry violation~\cite{Seiberg:1999vs}.  Recently, a new Lorentz symmetry violating 
theory of gravity, Ho\v{r}ava-Lifshitz Gravity~\cite{Horava:2009uw}, has been developed.
In this theory, gravity is power-counting renormalizable and unitarity is maintained. More
generally, it is possible to detect the hint of high energy physics from low energy 
effective field theory since the high energy phyiscs leads to a modification of the 
dispersion relations of different particles~\cite{Smolin:2003rk}. Motivated by this, 
there have been many searches for signs of Lorentz symmetry violating phenomena in 
terrestrial, astrophysical and cosmological settings
~\cite{AmelinoCamelia:1997gz,Bear:2000cd,Gleiser:2001rm,Sudarsky:2002ue,Jacobson:2002ye}, and the constraints on the magnitude of such phenomena are becoming better 
and better. 

The fact that different particles have different limiting speeds will open up the possibility of 
Cerenkov radiation for kinetic reasons, which is forbidden if Lorentz symmetry is kept. 
The most stringent constraint is from the highest energy cosmic rays. If Lorentz symmetry 
is broken, then there is no reason to keep the limiting speed the same in the dispersion 
relation of different particles. That the high energy cosmic rays which are possibly composed 
of hadrons, travelling astrophysical distances and times with  energy of about 
$3 \times 10^{11} GeV$ gives the following constraint on the limiting speeds of protons $c_p$
and photons $c_{\gamma}$ ~\cite{Coleman:1998ti}:
\be \label{problem}
\frac{c_p-c_\gamma}{c_\gamma} < 10^{-23} \, .
\ee
Even if the two limiting speeds are the same at tree level, if there is no specific symmetry it is 
hard to keep them the same at all orders without fine-tuning.  Moreover, dimension three 
operators which break Lorentz invariance are also highly constrained~\cite{Colladay:1996iz}.
The requirement is that their coupling constants should be much less than the Lorentz violation scale from simple dimensional counting.

Here we introduce supersymmetry as the candidate to solve the naturalness problem 
mentioned above for theories with Lorentz symmetry violation. Supersymmetry is an 
extended symmetry of spactime. In general, the algebra of supersymmetry is
built on the Poincare algebra. But here we are considering theories with Lorentz 
symmetry violation and thus the initial symmetry of space-time is not the Poincare 
symmetry. In the context of Ho\v{r}ava-Lifshitz gravity the initial symmetry is
$SO(3)$, the group of spatial rotations. Here, we will construct a new supersymmetry 
algebra starting from $SO(3)$ rather than the Lorentz group $SO(3,1)$. 

In a supersymmetric theory the bosons and fermions which are in the same multiplet
will by symmetry have the same dispersion relation.  This will help us solve the naturalness
problem (\ref{problem}). Therefore, if all the particle are in the same multiplet, such as in 
$N=8$ supergravity or $N=4$ super-Yang-Mills theory, the naturalness problem 
(\ref{problem}) should be completely absent. If not all particles are in the same multiplet, 
for example in the Minimal Supersymmetric Standard Model (MSSM), a gauge symmetry 
in combination with supersymmetry can keep the limiting speed of all  particles the same. 

In \cite{GrootNibbelink:2004za,Bolokhov:2005cj,Berger:2001rm}, the authors studied Lorentz violation in supersymmetric models in which the superalgebra itself did not violate the Lorentz 
symmetry, which is motivated from the low energy effective field theory point of view. 
At high energy, the superalgebra needs to be changed because the spacetime symmetry
explicitly breaks Lorentz invariance. To study this problem, we need to
work in the context of a specific theory with Lorentz violation at high energy, which 
is the reason that we here use Ho\v{r}ava-Lifshitz theory to construct supersymmetry. 
This theory is complete in the ultraviolet (UV), and in the infrared (IR) the Lorentz symmetry 
is emergent. The supersymmetry constructed here can be extended to other Lorentz 
violating theories.  

We shall start with the spacetime symmetry and scaling symmetry of the Lorentz violating 
Lifshitz theory in Sec.~\ref{symmetry}. We derive the supersymmetry generators and their
commutators in Sec.~\ref{lagrangian} from the free field Lagrangian. Then we will use
superspace language to construct the susy langrangian and interactions in 
Sec.~\ref{superspace}. So far, the discussion does not contain gauge fields. By 
introducing a gauge symmetry in Sec.~\ref{gauge}, we can ensure that different particles in
different multiplets also have the same limiting speed, which solve the naturalness problem. 
In Sec.~\ref{break}, we will discuss the relationship between susy breaking and Lorentz 
violation, and we will give the conclusion in the  last section.

\section{$SO(3)$ $\times$ Scaling Symmetry}
\label{symmetry}

Lifshitz theory explicitly breaks Lorentz invariance, and the remaining
symmetry is  $SO(3) \times$ Scaling Symmetry, spatial rotations
and the anisotropic scaling between space and time introduced in
\cite{Horava:2009uw}. With these symmetries, there 
is no reason that the limiting velocities of all species be the same.
Demanding agreement with the cosmic ray experiments would require
a fine-tuning of the parameters in the model to unbelievable accuracy. 
In order to avoid having to do this fine-tuning, we try to extend the $SO(3)$ 
symmetry by adding supersymmetry.

The Lorentz symmetry $ISO(3,1)$ algebra is composed of commutators of the translation generator $P_\mu$ and the spactime rotation generator $M_{\mu \nu}$,
\be
[P_\mu,P_\nu]=0
\ee
\be
[M_{\mu\nu} , P_\lambda]=i \left(   \eta_{\nu \lambda} P_\mu -\eta_{\mu \lambda} P_{\nu}   \right)
\ee
\be
[M_{\mu\nu}, M_{\lambda \sigma}]=
 i\left( \eta_{\nu\lambda}M_{\mu \sigma} +\eta_{\mu \sigma}M_{\nu\lambda} -
 \eta_{\mu \lambda}M_{\nu \sigma}-\eta_{\nu \sigma}M_{\mu \lambda}\right) \ .
\ee
To introduce supersymmetry, we add two-component Weyl spinor generator 
$Q_\alpha$ and $\bar{Q}_{\dot\alpha}$ to compose 
the extended algebra,
\be
[M_{\mu \nu}, Q_{\alpha}]= -i\left(\sigma_{\mu \nu}\right)_\alpha^{\ \beta} Q_\beta
\ee 
\be
[M_{\mu \nu}, \bar{Q}^{\dot\alpha}]= -i\left(\bar\sigma_{\mu \nu}\right)^{\dot\alpha}_{\ \dot\beta} \bar Q^{\dot\beta}
\ee 
\be
[P_\mu,Q_{\alpha}]=[P_\mu,\bar{Q}_{\dot\alpha}]=0
\ee
\be
\left\{ Q_{\alpha}, Q_{\beta}  \right\} =\left\{ \bar{Q}_{\dot\alpha},\bar{Q}_{\dot\beta}  \right\}=0
\ee
\be
\left\{ Q_{\alpha},\bar{Q}_{\dot\beta} \right\}=2 \sigma^\mu_{\alpha \dot \beta} P_{\mu} \ . \label{susyalgebra}
\ee

In our case we do not have boost invariance. Hence, we set the boost generator 
\be 
M_{0i}=0  \ ,
\ee 
which reduces the spacetime symmetry  to $SO(3)$. The symmetry group $SO(3)$ is 
equivalent to $SU(2)$, while  $SO(3,1)$ is equivalent to $SU(2) \times SU(2)$. 
To obtain a supersymmetric generator algebra in the case of  SO(3) the two spinors can 
be identified as
\be
Q_{\alpha}^{*} = \bar Q_{\dot \alpha} \ .
\ee

The other symmetry that we mentioned in Lifshitz theory is scaling symmetry, which posits 
that the Lagrangian is invariant under the following anisotropic scaling of space and time: 
\be
\textbf{x} \rightarrow e^{-\Omega} \textbf{x} \ , \ \ \ \ \ \ t \rightarrow  e^{- z \Omega} t \ ,
\ee
where z is taken to be an integer chosen as $z=3$ to obtain a theory of gravity in 4 spacetime dimension.
If we consider the momentum conjugate of the spacetime coordinate, the relationship between energy and momentum is as follows,
\be
(\Delta p)^{z} \sim \Delta E \ ,
\ee
where $p$ is the spatial momentum. This relationship will be reflected in the dispersion relation.

In the following section we will discuss how the scaling 
symmetry impacts on the supersymmetry from which we can derive the right form of the spinor generator and of the 
renormalizable supersymmetric Lagrangian. It will modify the susy algebra (\ref{susyalgebra}). 

\section{Supersymmetric Lagrangian}
\label{lagrangian}

In this section, we will from the most basic boson and fermion Lagrangian derive the 
explicit form of the spinor generator. 
The simplest free boson Lagrangian with two scalar fields is 
\be
\mathcal{L}_s= \partial_0 \phi^* \partial_0 \phi - \partial_i\partial^2\phi^*  \partial_i\partial^2\phi \ .
\ee 
As a first step, we neglected other kinetic terms which are super-renormalizable in the 
$z=3$ Lifshitz theory. These additional terms are important when discussing  
low-energy phenomenology. However, just from the simplified case taken above, 
we can clearly see the physics. Given the above bosonic Lagrangian, the
Lagrangian for the two Weyl fermions which are the supersymmetric partners
of the bosons is as follows
\be
\mathcal{L}_f= i \psi^{\dagger} \sigma^0 \partial_0 \psi +i\psi^{\dagger} \bar\sigma^i \partial_i \partial^2 \psi \ ,
\ee
where $\sigma^0$ is the identity matrix and $\bar\sigma^i$ is conjugate of $\sigma^i$ 
 
The susy transformation should change $\phi$ to $\psi$,
\be
\delta \phi = \epsilon \psi 
\ee
\be
\delta \phi^*= \epsilon^{\dagger} \psi^\dagger \ ,
\ee
while the transformation from fermions to bosons can be derived from 
the requirement that the symmetry is obeyed by the 
action. The result is
\be
\delta \psi =-i \epsilon^\dagger \partial_0 \phi -i \sigma^i  \epsilon^\dagger \partial_i \partial^2 \phi
\ee 
\be
\delta \psi^\dagger =i \epsilon \partial_0 \phi^* +i \epsilon\sigma^i   \partial_i \partial^2 \phi^* \ .
\ee 
The whole action is invariant under these transformation up to some total derivatives. 
The commutator of the supersymmetric spinor generators 
is derived from the commutator $\delta_1\delta_2 -\delta_2\delta_1$.
\be
\left\{ Q_{\alpha},\bar{Q}_{\dot\beta} \right\}= 2 \sigma^0 P_0 +2\sigma^i P^2 P_i  \ ,
\ee
where $P^2 = P^i P_i$.

If supersymmetry is not broken, it is easy to see that the speed of light of the two 
superpartners are the same, because the supersymmetric transformation 
fixes the coefficients in front of $p^3$ for fermions and of $p^6$ for bosons.

\section{Superspace and Superderivatives}
\label{superspace}

In this section, we will derive the superspace formulation of the modified supersymmetry proposed in the previous section. Because the dimension of the superfield is zero in the $z=3$ Lifshitz theory, the D-term and F-term in the theory are different from the case of regular supersymmetry. In the end, we will give the interaction Lagrangian which includes some new interactions between bosons and fermions.

Constructing supersymmetry can proceed by two methods, one is by direct construction 
as above, and the other is from the superspace formalism making use of general superfields.
In the following we illustrate this superfield method,  a method which makes it easy to see why 
the ``speed of light" is the same for different species, and which also makes it easy to 
deal with interactions. Below, $\theta$ denotes the superspace variable. 

A superfield transforms as follows under the action of the extended spacetime symmetry,
\begin{equation}
 S(x^\mu, \theta, \theta^\dagger ) \rightarrow \exp (i(\xi Q + \xi^\dagger Q^{\dagger} -a^{\mu}P_\mu)) S(x^\mu, \theta, \theta^\dagger ) 
\end{equation}
where the generators are 
\begin{equation}
 P_\mu=i\partial_\mu
\end{equation}

\begin{equation}
 i Q_\alpha = \frac{\partial}{\partial \theta^\alpha} - i\hat{\sigma}^\mu_{\alpha \beta} \theta^{\dagger \beta} \partial_\mu
\end{equation}

\begin{equation}
 i Q^\dagger_\beta = \frac{\partial}{\partial \theta^{\dagger\alpha}} - i\theta^{ \alpha}  \hat\sigma^\mu_{\alpha \beta}  \partial_\mu \ .
\end{equation}
$\hat\sigma$ is a newly defined sigma matrix in the Lifshitz theory,
\begin{equation}
 \hat\sigma^0 = \sigma^0 
\end{equation}
\begin{equation}
 \hat\sigma^i = \sigma^i f(P) \ .
\end{equation}
The information about the scaling symmetry of Lifshitz theory is hiding in the redefined sigma matrix $\hat\sigma^\mu$. After the redefinition, the susy algebra and superfield are written in the same form as the regular supersymmetry.
$f(P)$ is a function of the magnitude of the spatial momentum. In the $z=2$ case, the 
spinor generator algebra is simpler, $\{Q_\alpha, \bar {Q}_\beta\}= 2 P^2$. In Sec.~\ref{lagrangian}, we considered the marginal kinetic operator, $f(p)=P^2$. 
Including some super-renormalizable kinetic operators, $f(P)=1+P^2$, and in the low energy limit, the supersymmetry algebra 
returns to the algebra with Lorentz symmetry.

From the spinor generator algebra,
\begin{equation}
 \left\{ Q_{\alpha},{Q}^\dagger_{\beta} \right\}= 2 \hat\sigma^\mu P_\mu \ ,
\end{equation}
we can derive  the fermionic derivatives $D_\alpha$ and  $D^\dagger_\beta$, which anticommute 
with the supersymmetric generator, and are written as follows
\begin{equation}
 D_\alpha= \frac{\partial}{\partial \theta^\alpha} +i\hat{\sigma}^\mu_{\alpha \beta} \theta^{\dagger \beta} \partial_\mu
\end{equation}
\begin{equation}
 D^\dagger_\beta = -\frac{\partial}{\partial \theta^{\dagger\alpha}} - i\theta^{ \alpha}  \hat\sigma^\mu_{\alpha \beta}  \partial_\mu \ .
\end{equation}

Applying the superderivatives as a constraint, the chiral superfield $\Phi$ can be defined, which
satisfies
\begin{equation}
 D^\dagger_\beta \Phi=0 \ .
\end{equation}
Almost all the formulas are the same as in the case of regular susy if we use the 
redefined $\hat\sigma$. In particular, the superfield can be expanded in superspace as
\bqa
 \Phi(x^\mu,\theta,\theta^\dagger) &=& \phi +\sqrt{2} \theta \psi +\theta \theta F + i \theta \hat\sigma^\mu \theta^\dagger \partial_\mu \phi + \frac{i}{\sqrt{2}}
\theta \theta \theta^\dagger \hat{\bar \sigma}^\mu\partial_\mu \psi   \nonumber\\
&& -\frac{1}{4} (\partial_0\partial^0 + \partial_i\partial^i P^4) \phi \theta\theta \theta^\dagger \theta^\dagger \ ,
\eqa
where
\begin{equation}
 \partial_0\partial^0 + \partial_i\partial^i P^4 \equiv \hat \partial_\mu \hat \partial^\mu
\end{equation}
and where $F$ is the auxiliary field.

The dimension of all the operators in the superfield can be derived by power counting,
\begin{center}
\begin{tabular*}{0.5\textwidth}{@{\extracolsep{\fill}} c|cccc}
  &  $\phi$ and $\Phi$    &   $\psi$  &  $\theta$   & $F$  \\
	\hline
dim  & $\frac{3-z}{2}$ & $\frac{3}{2}$ &   $-\frac{z}{2}$& $z$  \\
\end{tabular*}
\end{center}
The kinetic term comes from the D term of $\Phi^\dagger \Phi$
\footnote{Please note the different partial derivative operators for fermions and bosons 
in the formula, one of which has a hat, while the other one does not have a hat.  }
\bqa
 \left[ \Phi^{\dagger}_i \Phi_j \right]_D &=& \left[ \Phi^{\dagger}_i \Phi_j \right]_{coeff. of \theta^\dagger \theta^\dagger \theta \theta}\nonumber\\
 &=& F_i^\dagger F_j +\frac{1}{2} \hat\partial_\mu \phi^\dagger_i \hat\partial^\mu\phi_j - \frac{1}{4}  \phi^\dagger_i \hat\partial_\mu \hat\partial^\mu\phi_j
- \frac{1}{4} \hat\partial_\mu \hat\partial^\mu \phi^\dagger_i \phi_j \nonumber\\
&&+ \frac{i}{2} \psi_i^\dagger \hat{\bar \sigma}^\mu \partial_\mu \psi_j +\frac{i}{2} \partial_\mu\psi_i^\dagger \hat{\bar \sigma}^\mu  \psi_j  \ .
\eqa

There are new terms in the Lagrangian of the form,
\be
\mathcal{L}_{NEW}=   [(\Phi^\dagger \Phi)^n ]_D \ ,
\ee 
where $n$ is an integer, and in the limit of low energies, the smaller $n$ terms 
dominate, since the scale of Lorentz violation suppresses these terms by a factor 
$(\frac{\phi}{M})^n$. And there are other possible terms $[\Phi^\dagger \Phi \Phi^n]_D + h.c.$ in the lagrangian which are forbidden by $U(1)$ or other symmetries.

The superpotential is constructed by holomorphic function of superfield,
\be
W(\Phi)=\sum_{i=1}^{\infty} g_n \Phi^n
\ee
and yields the interaction Lagrangian,
\be
\mathcal{L}_{int}=\int \d^4 x \left[ W \right]_F \ .
\ee
\be
\mathcal{L}_{int} = - \left(\frac{\partial W(\phi)}{\partial \phi_i}\right) \cdot  F - \frac{1}{2} \frac{\partial W(\phi)}
{\partial \phi_i \phi_j} \psi_i \psi_j + h.c. \, .
\ee
The first term includes $\left(\frac{\partial W(\phi)}{\partial \phi_i}\right)^2$, since 
there is a $F^2$ term from the kinetical D-term. However, in Lifshitz theory, we introduce 
many new D-terms which contain other terms depending on $F$, for example, the term 
$[\lambda(\Phi^\dagger \Phi)^2]_D$. There are no four fermion terms from the F-term due to the fact that $\theta^4=0$, 
but the newly introduced D term will give a four-fermion interaction. We write down the terms having F and four fermion interaction,
\be
\left[ \lambda (\Phi^\dagger \Phi)^2 \right]_D  \supset \lambda (2 \phi^\dagger \phi F^\dagger F  - 2 F^\dagger \psi \psi \phi ^\dagger-2 F \bar \psi \bar \psi \phi + \lambda \bar \psi \bar \psi \psi \psi)
\ee
After some calculation, we obtain the interaction terms written without the auxiliary field $F$,
\be
\mathcal{L}_{int}=\frac{ - \left( \frac{\partial W}{ \partial \phi} - 2\lambda \phi \bar \psi \bar \psi \right) \left(\frac{\partial W^\dagger}{ \partial \phi^\dagger} - 2\lambda \phi^\dagger  \psi  \psi   \right)}{1+2\lambda \phi^\dagger \phi} + \lambda \bar \psi \bar \psi \psi \psi - (\frac{1}{2} \frac{\partial W(\phi)}
{\partial \phi_i \phi_j} \psi_i \psi_j+ h.c)
\ee
This potential could be from usual scalar field vacuum expectation value, and also 
possibly from fermion condensation. Also, from the interaction Lagrangian, there 
are new phenomena which follow from the low energy effective field theory and which depend 
on the form of the superpotential.

\section{Gauge field and Gaugino}
\label{gauge}

Bosons and fermions in the same supermultiplet have the same limiting speed because of
supersymmetry. It is essential to see whether we can arrange for particles in different 
multiplets to have identical limiting speed by introducing a gauge symmetry and relating
particles in different supermultiplets by gauge transformations. If two supermultiplets 
are totally decoupled, then there is no reason that keep the coefficients of the kinetic 
terms the same. 

First of all, consider the kinetic term of a chiral superfield which is derived from 
$\mathcal{L}_{kin}= \left[ \Phi^\dagger \Phi \right]_D$. The supersymmetry 
transformation $Q$ fixes the form of $D_\alpha$, and then the form of the chiral 
supermultiplets is the same across all multiplets, and therefore the limiting speed of the two chiral supermultiplets are the same. We will see the coefficients of 
vector supermultiplets are also the same as that of the chiral supermultiplet since 
the gauge symmetry relates the two types of supermultiplets. 
  
A general vector supermultiplet can be written as 
\bqa
 V(x,\theta,\theta^\dagger) &=& C +i \theta \chi -i \theta^\dagger \chi^\dagger +\frac{1}{2} i \theta \theta \left[ M + i N  \right] -\frac{1}{2} i \theta^\dagger \theta^\dagger [M-iN] \nonumber\\
 &&\theta \tilde \sigma_1^\mu \theta^\dagger V_\mu +i\theta \theta \theta^\dagger\left[ \lambda^\dagger +\frac{i}{2} \tilde{\bar\sigma}_2^\mu \partial_\mu 
  \chi \right] -i \theta^\dagger\theta^\dagger \theta \left[ \lambda + \frac{i}{2} \tilde{\sigma}_2^\mu \partial_\mu 
  \chi \right] \nonumber\\
&& +\frac{1}{2}\theta\theta \theta^\dagger\theta^\dagger \left[ D-\frac{1}{2} \hat\partial_\mu \hat\partial^\mu C \right] \ ,
\eqa
where $\tilde{\sigma}_{1 or 2}$ is some newly defined sigma matrix  which only obeys the $SO(3)$ 
spatial symmetry. And $C$, $M$, $N$, $D$ are scalar fields, $\chi$, $\lambda$ are Weyl spinor, and $V_\mu$ is a vector field. 
A gauge transformation leads to the following change in the form of a superfield: 
\begin{equation}
 V(x,\theta,\theta^\dagger) \rightarrow  V(x,\theta,\theta^\dagger) + i(\Phi(x,\theta, \theta^\dagger)-\Phi^\dagger(x,\theta,\theta^\dagger) ) \ ,
\end{equation}
where 
\bqa
 \Phi(x^\mu,\theta,\theta^\dagger)-\Phi^\dagger(x,\theta,\theta^\dagger)  &=& (\phi-\phi^\dagger) +\sqrt{2} (\theta \psi- \theta^\dagger \bar \psi) +\theta \theta( F -F^\dagger)+ i  \theta \hat\sigma^\mu \theta^\dagger \partial_\mu (\phi-\phi^\dagger)  \nonumber\\&& + \frac{i}{\sqrt{2}}
\theta \theta \theta^\dagger \hat {\bar \sigma}^\mu \partial_\mu \psi -\frac{i}{\sqrt{2}}
\theta^\dagger \theta^\dagger \theta \hat { \sigma}^\mu \partial_\mu \bar \psi 
 -\frac{1}{4} \theta\theta \theta^\dagger \theta^\dagger \hat \partial_\mu \hat \partial^\mu (\phi-\phi^\dagger) 
\eqa
The way to construct a gauge symmetry is to introduce a chiral superfield, by which 
the sigma term is fixed, $\tilde{\sigma} =\hat{\sigma}$. 
Therefore, if the gauge field is coupled to the chiral supermultiplet, the coefficients 
of the kinetic terms are the same for all particles.

The component $V_\mu$ in the vector supermultiplet is a regular gauge field, and the field strength is derived from the field strength superfield $W_\alpha = D^{\dagger2}D_\alpha V$ ,
\be
W_\alpha (y,\theta) =i \lambda_\alpha - \left[  \delta_\alpha^\beta D(y) +\frac{1}{2}i(\hat \sigma^\mu \hat {\bar\sigma}^\nu)_\alpha^\beta V_{\mu\nu} (y) \right]\theta_\beta
+ \theta^2 \hat\sigma^\mu_{\alpha\beta} \partial \lambda^{\dagger \beta}
\ee
where $V_{\mu\nu}$ is the field strength
\be 
V_{\mu\nu} = \partial_{\mu} V_\nu - \partial_\nu V_\mu \ .
\ee
The Lagrangian of the Super-Yang-Mills field is the F-term of $W_\alpha W^\alpha$,
\bqa
  \mathcal L &=& \frac{1}{4} \left[ W_\alpha W^{\alpha} \right]_F +h.c. \nonumber\\
  &=& -\frac{1}{4}\hat V ^{\mu\nu}\hat V_{\mu\nu} +i\lambda \hat \sigma^\mu \partial_\mu \lambda^\dagger -\frac{1}{4} \hat V ^{\mu\nu} \left( * \hat V_{\mu\nu}\right)  \ ,
\eqa
where the redefined field strength is 
\be 
\hat V_{\mu\nu} = \hat \partial_{\mu} V_\nu - \hat \partial_\nu V_\mu \ ,
\ee
and all the derivative terms in the chiral field change from being normal partial 
derivatives to gauge covariant derivatives, but note that $\hat \sigma^{\mu}$ does not change. 
The covariant kinetic term of chiral superfield is as follows, 
\be
\int d^4\theta \Phi^\dagger \e {gV}\Phi \ .
\ee

The limiting speed of the gauge field is same as that of the gaugino, and also 
the same as that of a chiral field. 

Note that our construction of the gauge theory is different from the suggestion of 
~\cite{Chen:2009ka}. In order to make the Lagrangian gauge invariant, it follows
from the gauge transformation of a derivative term of a matter field that we need 
introduce the gauge field with the infinitesimal transformation rule,
\be
\delta A_\mu = \hat \partial_\mu  \epsilon + i f \epsilon A_\mu \ ,
\ee
where $f$ is the structure constant. In the Abelian gauge field case, $f=0$.

\section{SUSY Breaking}
\label{break}

The $N=0$ Lagrangian contains terms such as $\partial_i^2 \phi \phi^4$, which is 
renormalizable in the $z=3$ case, so the loop corrections will make the speed of light 
run with the scale. In the supersymmetric Lifshitz theory, there is the same interaction 
term containing derivatives in this way. This will lead to a quantum correction of 
the speed of light, but the supersymmetry will cancel the loop effect between bosons and 
fermions. Hence, this renders the theory free of the need of fine-tuning the speed of 
light of different particles. 

If SUSY is broken, there is a soft supersymmetry breaking term introduced which make 
the speed of light of particles different. One way to keep the difference small or absent is 
to break supersymmetry at a low energy scale (Lorentz symmetry is emergent which is equivalent to $z=1$ Lifshitz theory), and then the correction to the speed of light is  
not a soft term any more, and cannot yield different corrections to fermions and bosons 
from the Hidden Sector. 
On the other hand, if supersymmetry breaks around or above the scale of Lorentz symmetry
violation, the speed of light becomes related to a soft term, which makes it suffer 
from the fine-tuning problem to obtain agreement with the results from the cosmic ray 
experiments. Therefore, we obtain the constraint that the susy breaking scale 
should be smaller than the Lorentz violation scale.

\section{Discussion}
\label{dis}

In this paper, we have constructed an $N=1$ supersymmetric theory with scalar fields, 
gauge fields and fermion fields from a UV complete theory, the $z=3$ Lifshitz theory. 
The different particles have the same kinetic Lagrangian and hence the same limiting 
speed, which can solve the naturalness problem in theories with Lorentz symmetry violation. 

Our starting point is a supersymmetry algebra which is different from what is presented in previous papers 
on supersymmetric Lorentz violating theory~\cite{GrootNibbelink:2004za,Belich:2003fa,Berger:2001rm}. 
However, in the low energy limit, the supersymmetry algebra will be the same. Concerning 
the dimension three Lorentz violating operators, they do not arise in the Lifshitz theory, 
and thus this is also true in the supersymmetric case.

Since supersymmetry will make the loop calculations much different from the 
$N=0$ field theory case, it is important to calculate the renormalization group and 
find  the fixed point of the Lifshitz theory in the supersymmetric framework. 
In the original paper on Ho\v{r}ava-Lifshitz theory~\cite{Horava:2009uw}, 
a power-counting renormalizable theory of gravity, the author pointed that in the UV, and in 
the $z=3$ case, there is a free-field fixed point. In \cite{Dhar:2009dx,Dhar:2009am}, 
the authors studied the $z=3$ Lifshitz theory with fermions and gauge fields and 
calculated its renormalization group. And in \cite{Iengo:2009ix}, the authors studied the one-loop renormalization of scalar field in Lifshitz theory.
By introducing supersymmetry, there should 
be some changes in the renormalization group equations and in the conclusions
concerning fixed points.

\section*{Acknowledgements}
We would like to thank 
Robert Brandenberger, Alessandro Cerioni, Keshav Dasgupta, Maxim Pospelov,Oriol Pujolas
for helpful discussions.

 \bibliography{lifshitzsusy.bib}
 \bibliographystyle{plain}

\end{document}